\documentclass[conference]{ieeeaccess}

\usepackage{cite}
\usepackage{amsmath,amssymb,amsfonts}
\usepackage{algorithmic}
\usepackage{graphicx}
\usepackage{textcomp}
\usepackage{xcolor}
\usepackage{subfigure}

\def\BibTeX{{\rm B\kern-.05em{\sc i\kern-.025em b}\kern-.08em
    T\kern-.1667em\lower.7ex\hbox{E}\kern-.125emX}}
\graphicspath{{figs/}}

\usepackage{caption} 
\makeatletter
\def\endthebibliography{%
  \def\@noitemerr{\@latex@warning{Empty `thebibliography' environment}}%
  \endlist
}
\makeatother

\usepackage{footmisc}

\begin{document}

\history{Date of publication xxxx 00, 0000, date of current version xxxx 00, 0000.}
\doi{}

\title{A converged architecture for processing 32 Tbps of physics data in real-time at the LHCb experiment}

\author{Roel~Aaij,
Christina~Agapopoulou,
Thomas~Boettcher,
Dorothea~vom~Bruch,
Daniel~Hugo~C\'ampora~P\'erez,
Adrian~Casais~Vidal,
Tommaso~Colombo,
Daniel~C.~Craik,
Tim~Evans,
Placido~Fernandez~Declara,
Marianna~Fontana,
Vladimir~V.~Gligorov,
Arthur~Hennequin,
Louis~Henry,
Brij~Kishor~Jashal,
Saverio~Mariani,
Rosen~Matev,
Niklas~Nolte,
Niko~Neufeld,
Arantza~Oyanguren,
Alberto~Perro,
Flavio~Pisani,
Renato~Quagliani,
Florian~Reiss,
Kate~A.~Richardson,
Alessandro~Scarabotto,
Rainer~Schwemmer,
Patrick~Spradlin,
Marian~Stahl and
Jiahui~Zhuo}


\address[]{For affiliations, see Acknowledgements section}


\corresp{Corresponding authors: Daniel Hugo Cámpora Pérez (e-mail: dcampora@cern.ch), Flavio Pisani (e-mail: flavio.pisani@cern.ch), Rainer Schwemmer (e-mail: rainer.schwemmer@cern.ch).}

\begin{abstract}

The LHCb detector at the Large Hadron Collider has been upgraded to acquire an unprecedented 32~Tbps of particle-collision data to provide new insights in the High Energy Physics domain. The data produced by the detector is filtered in real-time to select interesting collisions. As part of the upgrade, a pre-filtering stage has been removed leading to a factor~40 increase in data rate. To deal with the high throughput demands of LHCb real-time data processing, we present an off-the-shelf network architecture using zero-copy techniques in conjunction with an efficient, fully-GPU-based filter. Our converged architecture is able to process the full 32~Tbps of particle-collision data in real-time, the highest in any physics experiment to date. Our result extends the reach of the LHCb physics programme and sets a new standard for real-time data processing at particle physics experiments.

\end{abstract}


\titlepgskip=-21pt

\maketitle

\section{Overview of the problem}

The Standard Model of particle physics is a successful theory that describes nature with remarkable precision. To produce the very rare events that may bear hints of \emph{new physics}, the Large Hadron Collider (LHC)~\cite{LHC} accelerates and collides billions of particles per second. The results of these collisions are observed by the four main LHC particle detectors. The high data rates produced by LHC detectors require \emph{trigger} systems that filter the data by selecting only interesting collisions in real-time~\cite{fruhwirth_data_2000}. The effectiveness of the LHC experiments depends on excellent performance in all steps of the chain, from detection to storage. Trigger systems play a crucial role as only the data they select will be available for later analysis.

The LHC detectors have similar architectures for data acquisition (DAQ). Bunches of particles cross every 25~nanoseconds, and each crossing produces multiple collisions. This is commonly referred to as an \emph{event}. Typically, a system called \emph{low-level trigger} uses specialized sensors to quickly decide whether an event can be ignored. If not, data acquisition from the whole detector is initiated. Raw data is then received from the detector and distributed among the \emph{readout units}. The \emph{builder units} then collect and assemble a complete ``snapshot" of each event, making it available to \emph{filter units}, which select events of interest based on the physics program of the experiment. This filter is called \emph{high-level trigger} (HLT).

Filtering events requires reconstructing some physical characteristics of the particles observed by the detector, by interpreting its signals. In practice, \emph{event reconstruction} mostly consists in solving pattern recognition problems. These problems range from finding the trajectory of charged particles from the traces left throughout the detector~\cite{strandlie_track_2010}, to obtaining the energy deposits of particles in calorimeter detectors~\cite{canudas_graph_2022}, covering a wide variety of techniques to find and classify the particles involved in each collision.

The LHCb detector~\cite{collaboration_lhcb_2008}~\cite{LHCbCollaboration2015} at the LHC has undergone an upgrade from 2019 to 2022 that has significantly enhanced its physics reach~\cite{LHCbTriggerOnlineTDR:2014}. \textbf{This upgrade has led to an unprecedented data-acquisition throughput of 32~Tbps which must be processed in real-time during the data taking periods of the experiment.}

The upgraded LHCb detector does not have a traditional low-level trigger system. Instead, all of the circa 30~million events per second delivered by the LHC are acquired and made available to a software-based high-level trigger system. The first stage of this system, called High-Level Trigger~1 (HLT1), performs a partial reconstruction and selection over the incoming events. Each event is decoded and reconstructed in real-time, obtaining features that allow to classify and accept or discard the event. Without a pre-selection by a low-level trigger, the LHCb HLT1 handles the full rate of events observed by the detector, a novelty among LHC experiments. Removing the low-level pre-selection results in a better quality selection, capable of considering the entire detector instead of just specialized sensors, but as a consequence \textbf{the upgraded LHCb must acquire and filter 40 times more data than before}~\cite{LHCbComputingTDR:2018}.

In this paper, we present a dense system that implements the data collection, redistribution, assembly and first stage of trigger, using only 164~servers. Our system ingests data at 32~Tbps and applies a 30:1 filter upon physics reconstructed particles. We innovate in the field by developing a converged architecture using zero-copy techniques and a fully-GPU-based trigger. We show the scalability of our solution and evaluate its performance across a variety of metrics. We compare to the state-of-the-art and determine that our system is capable of processing the highest data rate in software in our domain. Finally, we elaborate on the success of the LHCb data acquisition and first trigger stage and its impact in the field.

\section{Current State of the Art}

\begin{table*}[t]
\centering
\resizebox{\textwidth}{!}{%
\begin{tabular}{lcccccc}
\hline
                                                                 & Year & \begin{tabular}[c]{@{}c@{}}Experiment\\ data rate\end{tabular} & \begin{tabular}[c]{@{}c@{}}Software input\\ data rate\end{tabular} & \begin{tabular}[c]{@{}c@{}}Event size\\ and rate\end{tabular} & \begin{tabular}[c]{@{}c@{}}Software rejection /\\ compression factor\end{tabular} & Considered system size                                                                                   \\ \hline
Cobalt~\cite{broekema_cobalt_2018,romein_lofar_2010}                                                           & 2014 & 240 Gbps                                                      & 240 Gbps                                                           & N/A                                                           & 3:1                                                                               & \begin{tabular}[c]{@{}c@{}}8 nodes\\ 16 Intel Xeon Gold 6140\\ 16 NVIDIA K10\end{tabular}                \\ \hline
CHIME~\cite{the_chime_collaboration_overview_2022}                                                            & 2017 & 13 Tbps                                                       & 6.5 Tbps                                                           & N/A                                                           & 200:1                                                                             & \begin{tabular}[c]{@{}c@{}}256 nodes\\ 512 Intel Xeon E5-2620v3\\ 1024 AMD FirePro S9300 X2\end{tabular} \\ \hline
Belle II~\cite{aggarwal_snowmass_2022}                                                         & 2019 & 16 Gbps                                                       & 8 Gbps                                                             & \begin{tabular}[c]{@{}c@{}}1.1 MB\\ at 30 kHz\end{tabular}    & 6:1                                                                               & \begin{tabular}[c]{@{}c@{}}100 nodes\\ 4\,800 CPU cores\end{tabular}                        \\ \hline
ATLAS Run 3~\cite{atlas_collaboration_operation_2020}                                                      & 2022 & 480 Tbps                                                      & 1.6 Tbps                                                           & \begin{tabular}[c]{@{}c@{}}1.5 MB\\ at 30 MHz\end{tabular}    & 100:1                                                                             & 40\,000 \emph{Processing Units}~\cite{atlas_collaboration_operation_2020}                                                                                   \\ \hline
CMS Run 3~\cite{Fontanesi:2842439}                                                        & 2022 & 480 Tbps                                                      & 1.6 Tbps                                                           & \begin{tabular}[c]{@{}c@{}}1.5 MB\\ at 30 MHz\end{tabular}    & 100:1                                                                             & \begin{tabular}[c]{@{}c@{}}200 nodes\\ 400 AMD Milan 64-core\\ 400 NVIDIA Tesla T4\end{tabular}          \\ \hline
ALICE Run 3~\cite{rohr_usage_2021,alice_collaboration_alice_2023}                                                      & 2022 & 28 Tbps                                                       & 7.2 Tbps                                                             & \begin{tabular}[c]{@{}c@{}}18 MB\\ at 50 kHz\end{tabular}   & 7:1                                                                               & \begin{tabular}[c]{@{}c@{}}250 nodes\\ 28\,908 CPU cores\\ 2\,800 GPUs\end{tabular}                    \\ \hline
\begin{tabular}[c]{@{}l@{}}LHCb Run 3\\ (this work)\end{tabular} & 2022 & 32 Tbps                                                       & 32 Tbps                                                            & \begin{tabular}[c]{@{}c@{}}150 kB\\ at 30 MHz\end{tabular}    & 30:1 (HLT1)                                                                       & \begin{tabular}[c]{@{}c@{}}164 nodes\\ 328 AMD EPYC 7502 32-core\\ 328 NVIDIA RTX A5000\end{tabular}     \\ \hline
\end{tabular}
}

\caption{Data acquisition systems of experimental physics applications.}
\label{tab:comparison-experiments}
\end{table*}

Physics experiments are custom-built as full system optimizations. The data acquisitions are prepared to process the design rate of their experiment. The opposite also occurs whereby constraints in data acquisition and triggering have an impact on the detector technologies. In addition, not only technology, but also price and energy consumption drive design decisions. We nevertheless establish a state-of-the-art based on the premise that they all are similar data reduction problems, where in particular the amount of data processed in low-level or high-level triggers is a free parameter where data acquisition and software innovation can make a difference.

In Table~\ref{tab:comparison-experiments} we show data acquisition figures for all experiments under consideration. The LHC experiments share a collision rate of 30~MHz of non-empty proton-proton collisions, and a collision rate of 50~kHz of non-empty Pb-Pb collisions. All detectors take data during either physics run, however ALICE~\cite{ALICE} is tuned for Pb-Pb data taking, and as such we consider the Pb-Pb regime for ALICE and the proton-proton regime for the other three LHC experiments.

ATLAS~\cite{ATLAS} and CMS~\cite{CMS} have similar data acquisition input data rates and event sizes. This is by design, as both are general purpose experiments with full detector coverage, being able to detect particles traversing the detector in any direction. Both ATLAS and CMS employ low-level trigger systems, built with custom electronics, that reduce the event rate by a factor~300 using information from calorimeters and specialized muon detectors~\cite{atlas_trigger,cms_collaboration_cms_2017}. ATLAS has a software high-level trigger executed on CPUs~\cite{atlas_hlt}. The CMS high-level trigger is instead partially accelerated with scientific GPU cards~\cite{cms_gpu_hlt}. Both high-level trigger systems reduce the event rate further by a factor~100.

The ALICE data acquisition system uses a hybrid approach. Part of the data is readout using a triggered architecture, while the majority of it is collected using a streaming DAQ paradigm. This second way of acquiring data consists of having a constant data-flow, which is then divided into constant time slices. Subsequently, the slices are compressed in hardware on the DAQ cards, significantly reducing the amount of data produced~\cite{Costa_2017}. Thanks to the bigger event size produced by Pb-Pb collisions, the events can be efficiently reconstructed on their GPU cluster. Their \emph{Time Projection Chamber} reconstruction dominates the software reconstruction, which allowed ALICE to offload part of its computation to accelerators sooner than other LHC experiments~\cite{rohr_alice_2012}. Several events from separate collision windows are gathered in \emph{time frames} to further saturate the GPU.

In High Energy Physics and beyond the LHC experiments, we consider Belle II~\cite{kou_belle_2018}. The experiment started data-taking in 2019 and is taking 1.1~MB events at a rate of 30~kHz, using a two-stage software high-level trigger that reduces the data rate 6:1. Belle II uses a CPU-only software trigger that parallelizes over events, discarding background events using a full reconstruction.

We also look at other physics experiments with streaming problems that are solved at a high rate. In radio astronomy, correlation and beam forming are two applications that usually go hand in hand and require dedicated signal digital systems to process and compress the data in real-time. The correlator is used for imaging modes that result in sky-images, while beamforming is used for time series analysis searching for patterns indicative of astronomical events such as pulsars or bursts. In contrast with High Energy Physics, event boundaries are not known in advance, since the data from images can range in size from bytes to terabytes of information. Similarly, event rates are also unknown.

Two high throughput data acquisition systems for radio astronomy are considered, Cobalt and CHIME. Cobalt implements both a Fourier transform and a cross-correlation stage in their GPU application, hence taking in all the DAQ input rate of 240~Gbps into their application. In contrast, CHIME implements the Fourier transform in a pre-processing stage in FPGA, resulting in an input rate reduced by half to the software-based correlator. The resulting compression rates can attain up to 3:1 and 200:1 in each respective experiment.

\newpage
\section{Innovations realized}

\subsection{Summary of innovations}

Our main contribution is a converged architecture that is dense and scalable to perform the real-time data processing of the 32~Tbps of physics collision data at the LHCb experiment, the highest data rate processed in software in any physics experiment. This is achieved by implementing the first stage of trigger as a GPU-based application with optimized implementations of reconstruction and selection algorithms.

\subsection{DAQ innovation}
\label{sec:daq-innovation}

Data acquisition in high energy physics is primarily a data collation problem. From a physics point of view, each event is a separate problem and can be processed independently. However, large particle detectors consist of a myriad of independent sensors, each monitoring a tiny fraction of the space near the collision point. Therefore, data acquisition systems receive time-ordered streams of data fragments from many sources (about 10\,000 in the case of LHCb) and must assemble these fragments into coherent snapshots of each event. This process is known as \textit{event building}.

The traditional approach to dealing with this complexity has two prongs:
\begin{itemize}
    \item  Low-level trigger systems: A low-level trigger, typically implemented on custom ASICs or FPGAs, collects only the signals from specialized sensors and uses only that information to decide whether or not it is worthwhile to acquire data from the rest of the detector. Thus, the low-level trigger handles a smaller amount of data, while also making the rest of the data acquisition system's job easier. However, because they don't acquire data from the whole detector and because they are implemented in custom hardware, low-level triggers are quite inflexible, and adapting them to changing experimental requirements is often not feasible.
    \item Deep-buffered networks: The data fragments corresponding to an event accepted by the low-level trigger must be collected in a single computer to be made available to the high-level trigger. On a computer network, this results in a many-to-one communication pattern. Many-to-one communications are subject to a well-known pathology called \textit{incast}, which causes throughput collapse “when a client simultaneously receives a short burst of data from multiple sources”~\cite{incast_original}. The most widely adopted solution in data-acquisition networks is to use network switches with buffers large enough to absorb the bursts~\cite{incast_report}. However, as network speeds leapfrog memory speeds, deep-buffered high-bandwidth switches are becoming a thing of the past.
\end{itemize}

Our innovation for the current generation of the LHCb data acquisition system is to take advantage of advances in HPC networking technology to build a compact data-acquisition system that can handle 32~Tbps of data using as few custom electronics as possible. In detail, we present the following DAQ innovations:

\subsubsection{Commercial-off-the-shelf components}
An FPGA-based PCIe card that receives data from up to 48~data sources in the detector had to be developed in-house~\cite{Cachemiche:2016reb}. No existing product could read that many optical links in a single card. With this single exception, which allows the DAQ system to be exceptionally compact, the system is built entirely from commercial-off-the-shelf components.

\subsubsection{Better buffering}
The FPGA-based data receiver cards are placed directly into the servers that perform the event building. Host memory is comparatively cheap compared to FPGA memory, and copious amounts can now be used to temporarily buffer the data-fragment streams. With this change, buffer capacities have increased from microseconds to seconds. These large source buffers enable advanced traffic scheduling, which can be used to ensure efficient network utilization and avoid incast altogether, eliminating the need for deep-buffer switches.

\subsubsection{Better network engineering}
We design our system topology to fit our problem needs. Our network is a fat-tree network with endpoints acting as both read and build units. This reduces the problem of event building to a well-known collective communication pattern: the all-to-all personalized exchange. Although developed for HPC applications, optimized scheduling and routing algorithms for all-to-all exchange~\cite{ftree_a2a} can be applied to event building. The LHCb DAQ system is the first physics experiment DAQ to exploit this possibility.

\subsubsection{Convergence}
We host the GPU cards that run the high-level trigger~1 in the same infrastructure that is required for hosting PCIe data-receiver and network interface cards, with minimal infrastructure cost overhead. This also allows us to take advantage of the zero-copy technology offered by HPC network cards and minimize memory bandwidth consumption by running the readout, building and filtering workloads on the same machine.

\begin{figure}[hbt!]
  \centering
  \includegraphics[width=\linewidth]{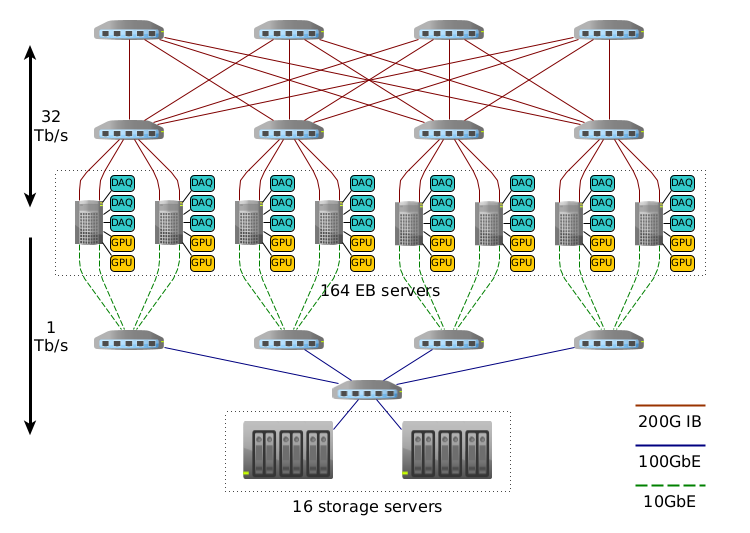}
  \caption{Overview of the current data-acquisition system of LHCb. The data from the detector is collected by the DAQ cards, subsequently it is collated using a dedicated InfiniBand network. The aggregated data is then filtered on the GPUs, and the selected fraction is sent to a storage system via an Ethernet-based network.}
  \label{fig:daq}
\end{figure}

Figure~\ref{fig:daq} shows a reduced-scale schematic representation of the final network architecture. Data arrives from the detector on the cards labelled DAQ. It is aggregated and buffered in the cluster node before being scheduled onto the HPC network (in red) for assembly in one of the other cluster nodes. The assembled data is sent through the GPUs for decoding and filtering. Accepted data is then sent out via the two onboard 10~Gb Ethernet ports to a storage layer for further, non-real-time processing.

\subsection{Algorithmic innovation}

To make effective use of the heterogeneous LHCb \emph{event builder} servers, our strategy is to implement the first stage of LHCb trigger software on GPUs. However, the comparatively small event size of up to 150~kB (see Table~\ref{tab:comparison-experiments}) and the fragmented processing requirements of each event pose a challenge for saturating many-core architectures, where the naive approach of offloading parts of the execution of each event would be dominated by transmission and synchronization. Instead, we present the following algorithmic innovations:

\subsubsection{Scheduling}

Our high-level trigger application comprises a sequence of algorithms that reconstruct the signals left throughout the detector for each event, with decision algorithms that may accept or reject events according to physics criteria. Events are physically independent and can be reconstructed in parallel.

We apply parallelism on three levels: 1) running several sequences in parallel, 2) batching multiple events in each sequence, and 3) exploiting data-parallel patterns in each reconstruction algorithm. We map these three levels of parallelism to GPU streams, blocks and threads respectively. The execution of each sequence is steered by a CPU thread and a GPU stream, such that several sequences can run in parallel without interfering or synchronising with one another.

Batching events introduces a scheduling problem. For a sequence with branching paths, such as conditional decisions to reject the event, each event might follow a differing path of execution~\cite{matev_configuration_2020}. The control flow of the batch of events could follow each of these possible paths, but that would result in many executions of the same algorithms with few events each time.

This problem has been solved by creating a \emph{multi-event scheduler} that considers the data and control flow constraints of the problem to create an execution schedule consisting of a plain sequence of algorithms. According to rules of conditional execution, algorithms receive the according execution mask that determines which events they act upon. Due to the nature of the application, this topo-sort problem has a large number of possible solutions. Creating the sequence of algorithms that is fastest to execute is an NP-complete problem~\cite{mertens_easiest_2003}.
To produce a reasonably efficient sequence that fits within the throughput budget, we implement heuristics guided by two principles that we found to be significant:
\begin{itemize}
\item{Spread data providers}: A more efficient pipeline is created by spreading algorithms with similar resource profiles. In this way, algorithms that are dominated by memory transfers and algorithms that are dominated by compute are interleaved, optimizing the overall resource usage.
\item{Expensive algorithms get tight masks}: Most algorithms only need to be run conditionally. Depending on the order in which they are executed, the knowledge of the outcome of previous algorithms can yield a tighter execution mask and reduce workload. We aim to schedule algorithms in a way that favors tighter event masks for more expensive algorithms.
\end{itemize}

Additionally to event parallelism,
we have exploited data parallelism within events according to the nature of each problem. Particle trajectories are independent, and thus we can seed and follow each trajectory in parallel. Locations where particles decay or collide are independent and therefore we can find them in parallel. The energy deposits in calorimeters have local maxima that can be reconstructed and assigned to particles independently. We have developed algorithms that have pushed the state of the art in each of their respective areas in physics reconstruction~\cite{CamporaPerez2021,placido2019parallel}.

\subsubsection{Memory}

The amount of global memory available on GPUs per core is scarce, orders of magnitude below what is available on CPUs per core, typically O(MB) in the former versus O(GB) in the latter. We account for this limitation by developing our application with well-defined boundaries for memory utilization as one of its core design principles.

The memory used by our application is statically configured at startup. Detector geometry constants are populated in global memory, available for the duration of the application execution. A double buffer is prepared for ingesting collision data efficiently, creating a pipeline of data collection and data processing.

Sequence execution is concurrently steered by a configurable number of threads. To forgo allocation overheads, we have developed a custom memory manager that allocates a configurable memory size on both host and device upon creation. Each sequence thread instantiates its own memory manager, which will manage the required data buffers during the entirety of the sequence execution.

We solve the memory limitation at the framework level. Our scheduler generates at startup a static list of data dependencies between algorithms in the sequence. Firstly, that allows us to automatically reserve and free data buffers on the go at the beginning of execution of each sequence step. And secondly, we allow developers to set buffer sizes but prevent them from directly interacting with the memory manager, such that we effectively prevent explicit dynamic memory allocations.

It is worth noting that the maximum amount of memory required by a sequence of algorithms is by definition unpredictable, given the unpredictability of physics collision data. The best one can hope for is to predictably fail and to treat these special cases. We achieve the former with statically configured memory sizes, and the latter by split-and-retry with smaller event batch sizes.

\subsubsection{Precision}

The quality of physics reconstruction is evaluated through a set of physics efficiency criteria, tested with Monte Carlo simulation data. The compartmentalized nature of the LHCb detector provides us with an opportunity to consider different precisions for each part of the detector.

Most of our application requires single precision. To fit the reconstructed trajectories over the entire detector we employ a Kalman filter~\cite{Kalman1960,CamporaPerez2018a}, which presents numerical instabilities when done in single precision for 1\% of the tested data. Instead, we apply the filter to a restricted region of the detector close to the interaction point. This simplification provides enough accuracy to cover the first stage of triggering, allowing us to use a single precision version of the fit.

Moving to half precision results in an unacceptable loss in physics efficiency at the permil-level in most of our reconstruction chain, with two exceptions in pattern recognition algorithms. We reconstruct the particle trajectories in the \emph{Vertex Locator subdetector}~\cite{LHCbVeloTDR:2013} using half precision for storage and single precision for arithmetic, and we reconstruct trajectory seeds in the \emph{Scintillating Fibers subdetector}~\cite{LHCbTrackerTDR:2014} using half precision for both storage and arithmetic. These two optimizations significantly reduce memory pressure in two of the hottest sections of our code.

Finally, we have developed our software using CUDA as a kernel language, and a flexible abstraction for API library calls~\cite{GPUTDR:2020,Aaij2019a}. We provide definitions for CUDA kernel syntax for HIP and CPU targets, and provide low-level optimizations in all three languages. We reach cross-architecture compatibility with a single source codebase. We have validated the precision of our algorithms to meet the quality requirements of the LHCb physics programme across architectures.

\section{How performance was measured}

\subsection{Data Acquisition System}

\begin{figure}[hbt!]
  \centering
  \includegraphics[width=\linewidth]{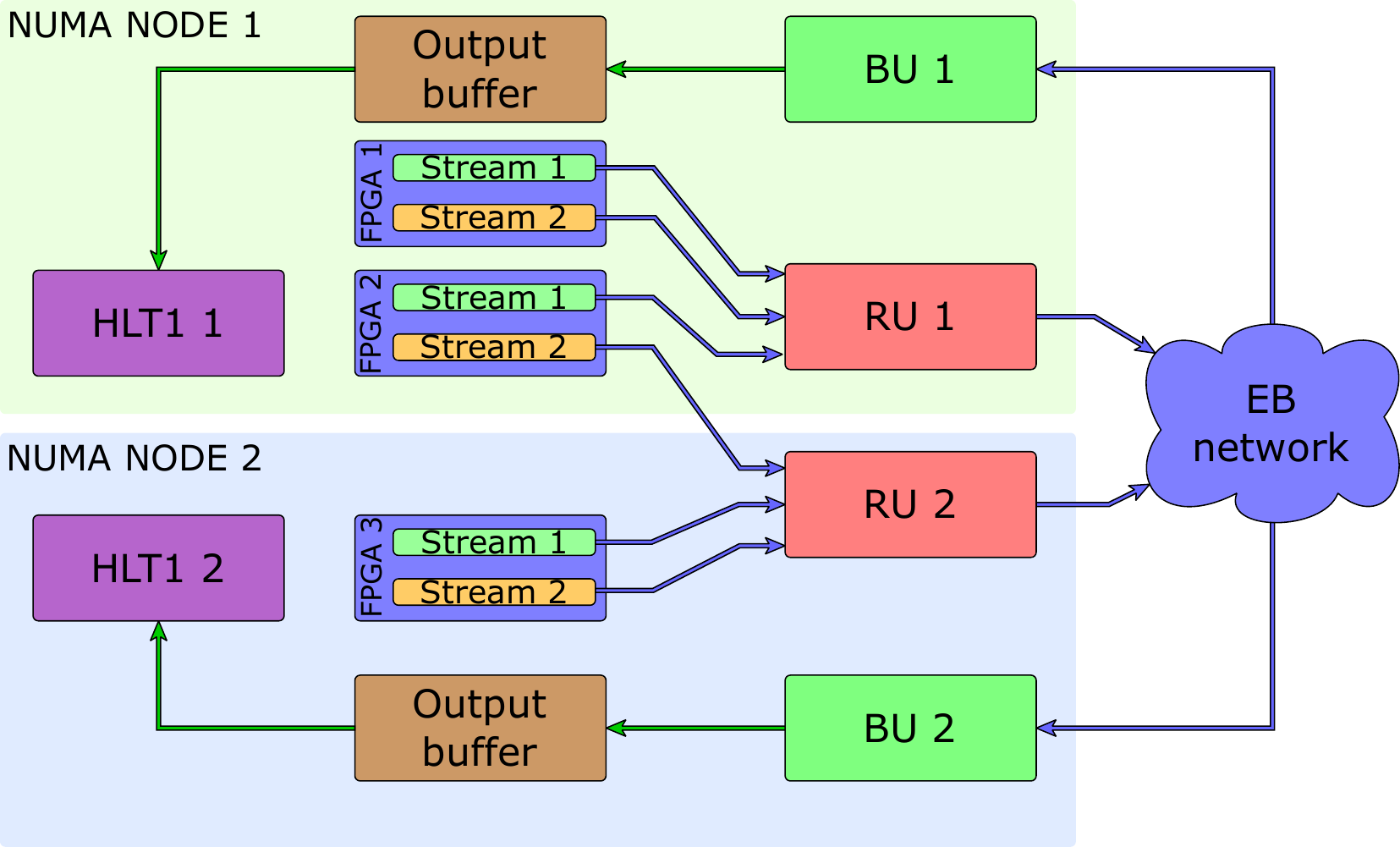}
  \caption{Data flow schema within one LHCb cluster node.}
  \label{fig:dataflow}
\end{figure}

In this section we provide a brief overview of the data acquisition system of the LHCb experiment~\cite{eb_rt_2022}. The data flow of an individual cluster node is shown in Figure~\ref{fig:dataflow}. Data arrives in the system via the FPGA data acquisition boards and their associated kernel drivers, and it is placed via DMA into a dedicated memory buffer handled by the device drivers. The Readout Unit (RU) application accesses the data via a user space API. From the RUs the data is then sent via a dedicated network~(cref.~\ref{sec:daq-innovation}) to the Builder Unit (BU) processes. Each BU receives and processes pieces of data, matching a different set of events, from all other cluster nodes. To minimise network congestion the application uses strict scheduling and it follows an all-to-all personalized exchange. Consequently the BU process sorts and collates the various event pieces and places them into a shared memory buffer, which acts as a queue between the Builder Unit process and the process orchestrating the GPU processing (HLT1). Data is then again copied via DMA into the GPU where it is decoded, reconstructed and filtered at a targeted 30:1 ratio. The remaining data, now greatly reduced in volume, is sent out through standard TCP sockets via the onboard NICs to the storage layer. The system is designed with a zero-copy paradigm and it takes advantage of the kernel bypass provided by DMA and RDMA technologies.

\begin{figure}[hbt!]
  \centering
  \includegraphics[width=\linewidth]{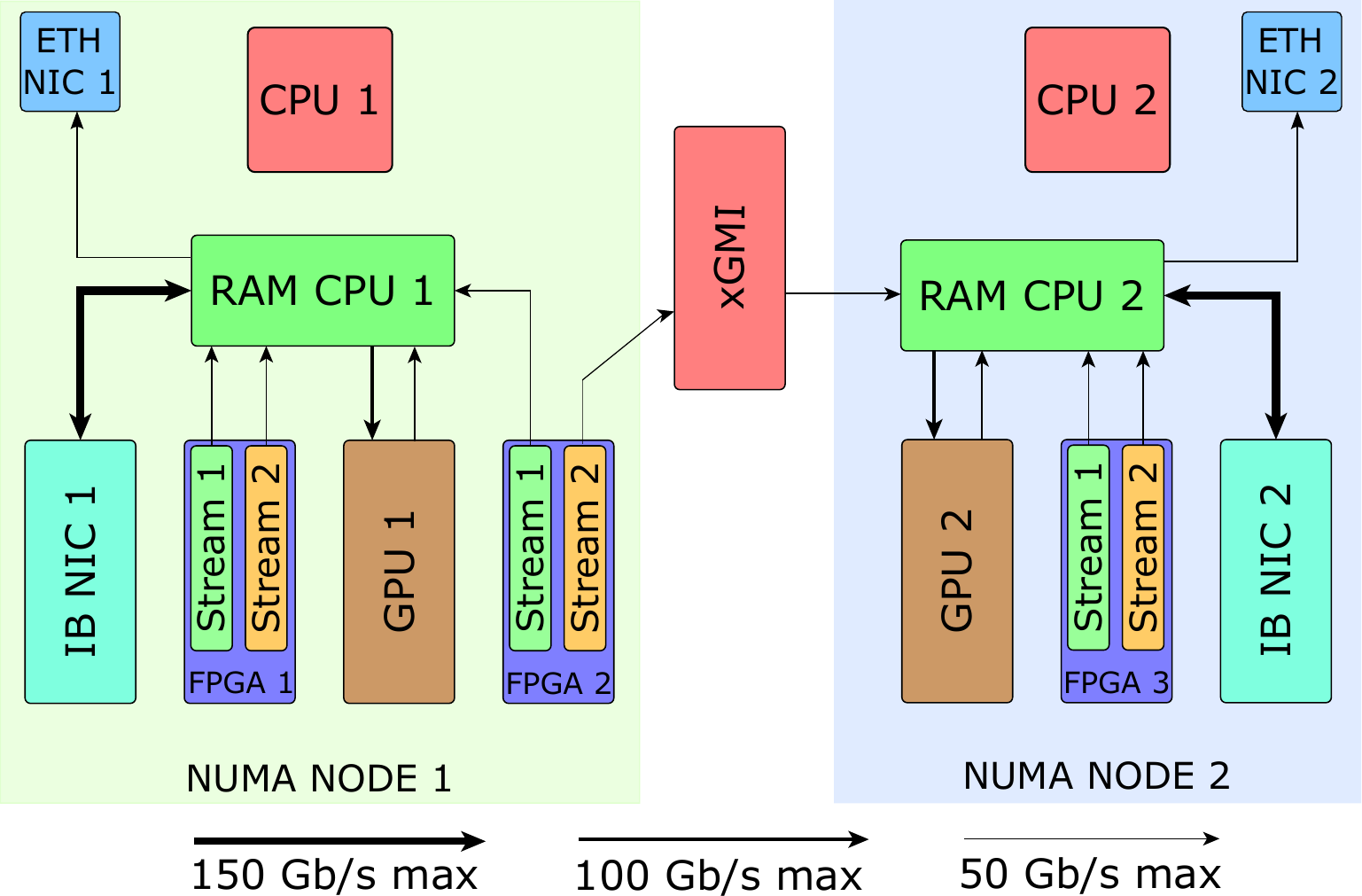}
  \caption{Hardware view of one LHCb cluster node.}
  \label{fig:single_server}
\end{figure}

The data flow matches the underlying hardware as shown in Figure~\ref{fig:single_server}. For two cards, data from the detector is received in the FPGA boards and sent to the memory node of the NUMA domain they are attached to. We tried to treat each socket of the node as independent entities as much as possible to avoid NUMA domain crossing effects, but could afford to put 3 FPGA boards per server. Every FPGA card consists of two independent sub sources, which allowed this to be done without negative impacts and helped to keep the system more compact. It also ensures that both network cards see, on average, the same amount of traffic. 

Since the FPGA cards are PCIe Gen~3 and the network cards are Gen~4, there is a natural limit to what the FPGAs can send which is only 3/4 of the network capacity -- i.e. 150~Gb/s. This leaves a comfortable operational margin for retransmissions and small scale congestion on the cluster network.  Data is then sent to the GPU. Only the selected data from the GPU is sent back to the server, which finally forwards it to the storage layer.

Computing nodes are interconnected via a non-blocking two-layer fat-tree~\cite{fat-trees} topology based on an InfiniBand HDR fabric. The full network is currently composed of 28 40-port HDR switches divided across the two layers of the fat tree, in particular the lower layer (leaf) counts 18 switches and the upper one (spine) counts 10. Each cluster node is connected using two independent network cards to one spine switch, and each leaf switch is connected to each spine via two links.

Contrary to a classical HPC application, which can usually be restarted from a previous checkpoint in case of problems, in our system any issue leads to a permanent loss of data. In particular memory stalls, hardware failures, application crashes and a plethora of other potential issues need to be carefully noted. For this reason the data flow is very closely monitored during data production at every step of the process to make it as efficient as possible, and to find out exactly where and when in the processing pipeline something goes wrong.

Subsequently all processes and hardware components are instrumented with performance counters, keeping track of how much data, fragments, time slices and other noteworthy objects have passed through and are currently in flight. As in the previous iteration of LHCb, this monitoring data goes into a central monitoring and control system, which consolidates the data for human consumption and aids in recovery procedures or recovers common problems automatically. During the commissioning phase, we use these counters to track performance and guide our optimisation efforts of the system. Throughput measurements presented here were gathered using this instrumentation.

\subsection{Platforms and Software Environment}

For this paper we perform our application benchmarks on our now completed cluster consisting of 164 nodes. Each node is equipped with two AMD EPYC 7502, 32-core CPUs and two NVIDIA RTX A5000 GPUs, each on one NUMA domain. The peak half/single precision performance considering only the GPUs in our cluster is 9~PFLOPS. We do not consider double precision as our application does not require it.

For the most part, our application executes on the GPU, however we must consider the boundary conditions of the entire system. Memory throughput must sustain the total rate of the entire DAQ chain of 500~Gbps on each NUMA domain. GPU memory throughput must have enough margin to sustain a data feed-in rate of 80~Gbps. The CPU utilization of the trigger application must not be too high in order not to interfere with other tasks of the data acquisition chain. Since we are dealing with a soft-real-time system, doing so would result in back-pressure and loss of data.

The software environment for our tests is shown in Table~\ref{tab:compiler-setup}. We use the STL thread implementation to achieve multi-threading, and zeroMQ to steer a finite-state machine that manages the state of the events under process.

\begin{table}[hbt!]
\centering
\begin{tabular}{ll}
\hline
Component          & Version  \\ \hline
Host compiler      & gcc 11.3.0   \\
Device compiler    & nvcc 12.1.66 \\
GNU libc           & 2.34 \\
OS                 & Red Hat 9.1  \\
Linux kernel       & 5.14.0-162.6.1 \\ \hline
\end{tabular}
\caption{Software environment}
\label{tab:compiler-setup}
\end{table}

We launch one application per GPU, with NUMA affinity set to match the CPU and memory node where the GPU resides. We connect to the event builder buffer (BU) and write to the HLT1 output buffer, both of the same affinity as the application.

\subsection{Measurements}

It is worth noting that our application is not bound by latency, but rather by throughput. The processing time of a single event only requires a latency guarantee insofar as it fits within the time budget affordable by the depth of the memory buffer. We have found this requirement to be relatively lax and therefore we do not report time-to-solution as part of our submission.

Reports generated by profilers generally are hard to apply to our throughput-oriented application. The optimizations proposed tend to yield faster kernels at the cost of a proportionally higher resource utilization, which results in lower throughput overall.

We collect the FLOPS of our software using the NVIDIA Nsight Compute profiler. We measure performance using the following metrics:

\begin{itemize}
  \item \textbf{Peak performance}, measured in $\frac{events}{second}$. We measure the maximum throughput achievable by our software application.
  \item \textbf{Sustained performance}, measured in $\frac{events}{second}$. The rate of the whole system including I/O has a well-defined target of 30~MHz, which is the design rate of non-empty proton-proton collisions at the LHC.
\end{itemize}

\section{Performance results}
This section is structured as follows. We first motivate some of the hardware choices behind the current LHCb cluster infrastructure. We then proceed to show multi server tests of our software application, where we present the scalability of our application. Finally, we discuss the sustained performance considering the most relevant boundary conditions of our system.

\subsection{Single server results}

\begin{figure}[hbt!]
  \includegraphics[width=\linewidth]{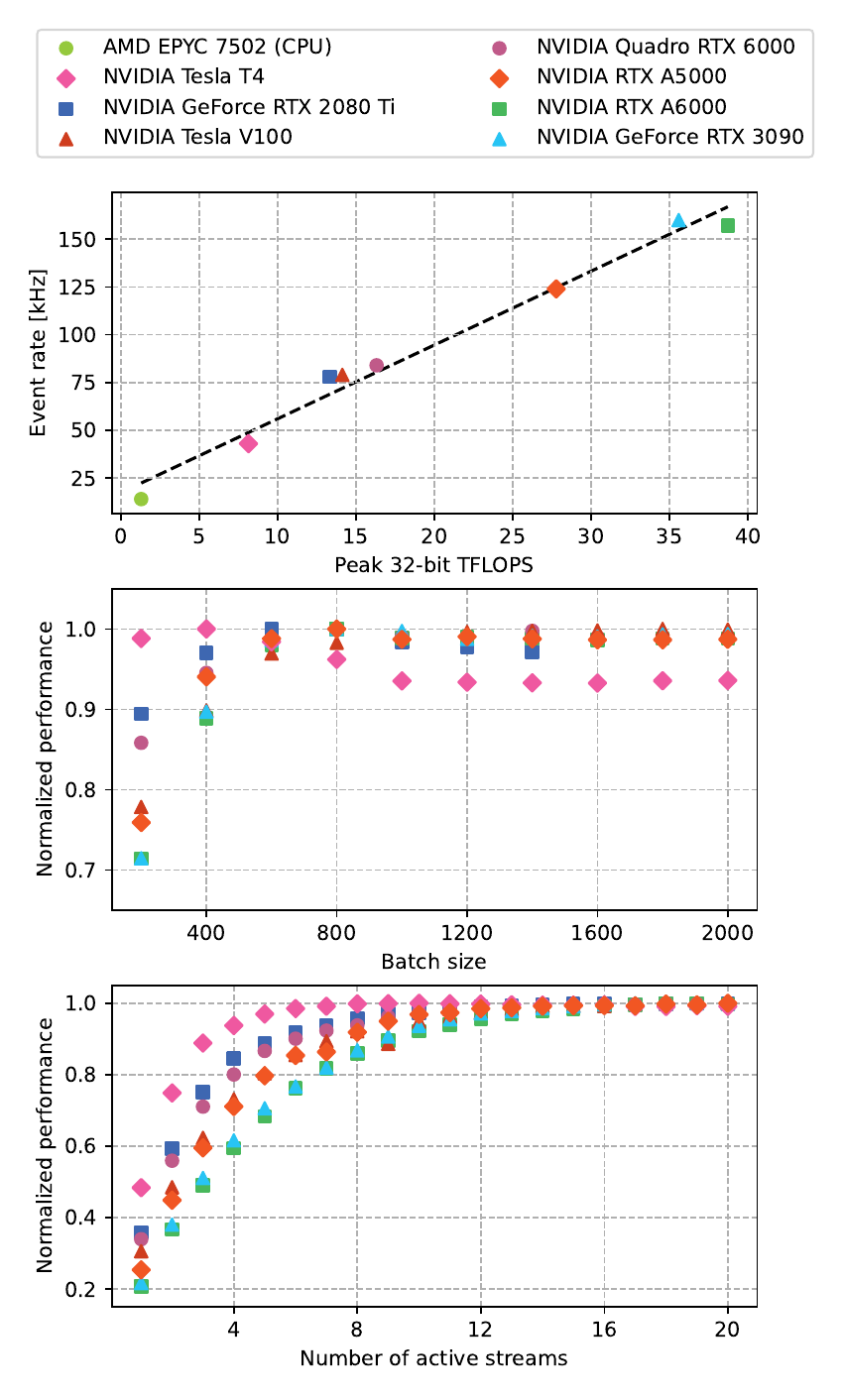}
  \caption{Top: Event rate attained by HLT1 across architectures. Middle: Performance of HLT1 as a function of batch size. Bottom: Performance of HLT1 with an increasing number of active streams (strong scaling).}
  \label{fig:ss_allen_scaling}
\end{figure}

Figure~\ref{fig:ss_allen_scaling} shows the scalability of our application for several GPU architectures and one CPU architecture. We consider consumer, professional and scientific cards. The top Figure shows the scaling of our application against the peak 32-bit FLOPS of the architectures under consideration. Our application scales linearly against peak 32-bit FLOPS.

A parameter scan of the batch size is shown in the middle Figure. The best performance is generally achieved for a batch size between 600 and 800~events. The bottom Figure shows the performance obtained as a function of the number of threads-streams in our application. We observe that more active streams are required to saturate higher-end cards. Performance levels out for all cards with a configuration of at least 16~active~streams.

\begin{figure}[hbt!]
  \centering
  \includegraphics[width=\linewidth]{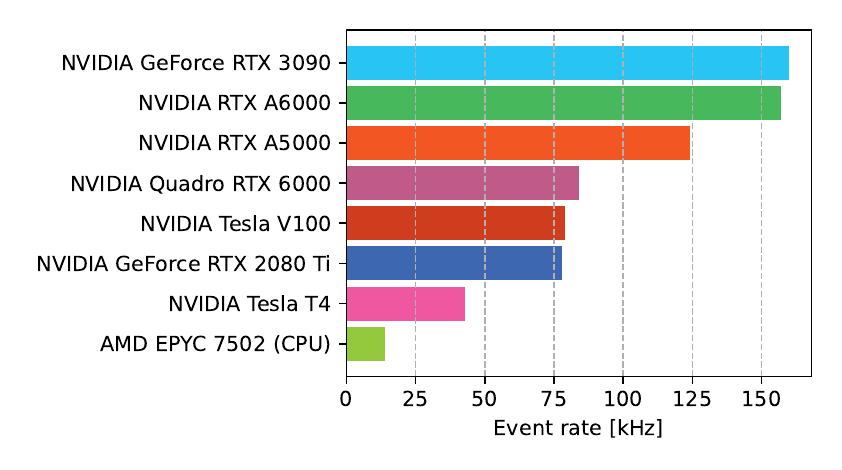}
  \caption{Throughput of LHCb HLT1 across architectures.}
  \label{fig:ss_allen_throughput}
\end{figure}

Figure~\ref{fig:ss_allen_throughput} shows the peak performance for the tested processors. The architecture choice for HLT1 took into consideration the price, throughput and power consumption between architectures~\cite{Aaij2021}. The single precision envelope of our application widened our hardware choice beyond the scientific GPU market, and the high-reliability requirements of our application moved us away from consumer cards. The network synchronicity of our system imposes the restriction that we install a symmetric system, equipping each node with the same number of GPUs. We decided to install two NVIDIA~RTX~A5000s on each node, which surpass our throughput requirement while remaining on-budget, and leaves open the possibility of installing an additional GPU in the future on the last remaining free PCIe~slot.

\subsection{Multi server results}

We have performed scalability tests for our cluster for the production physics sequence. \textbf{We report a peak performance of all GPU cards combined of 41~MHz in isolation, with 36\% headroom over the required event rate.} This result allows us to extend the physics reach of the experiment by considering additional processing as part of HLT1.

We obtain a combined performance of 57/82 TFLOPS in half/single precision and a MIPS of 5\,596\,800. Each GPU reports a DRAM utilization of 107/93 GBps for read and write respectively. We note that several algorithms in our application are memory bound, in particular all decoding algorithms and data preparation steps, which are an essential part of physics reconstruction sequences.

\begin{figure}[hbt!]
  \centering
  \includegraphics[width=\linewidth]{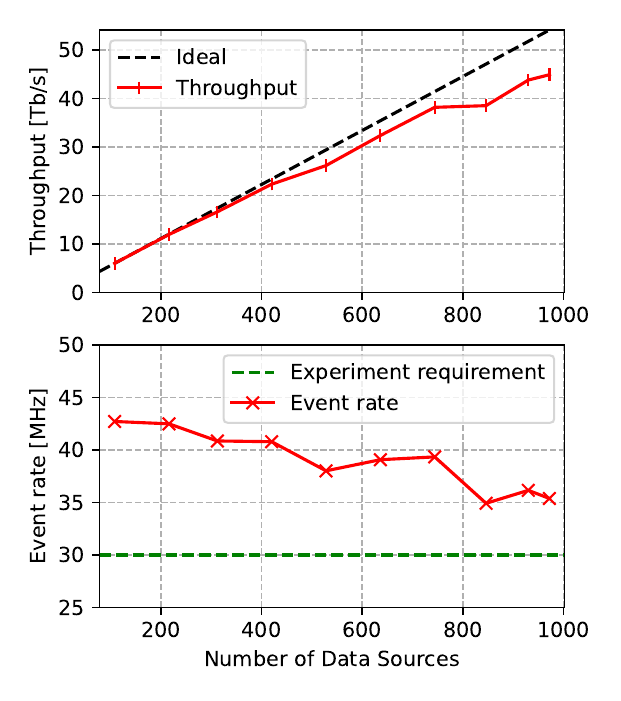}
  \caption{Scalability test of the event builder (EB). The top plot shows the throughput vs the number of network interfaces used and the ideal scaling of the application. The bottom plot shows the event rate vs the number of network interfaces and the experiment's requirement.}
  \label{fig:eb_scalabity}
\end{figure}

We show the weak-scaling for the event building application in Figure~\ref{fig:eb_scalabity}. The top figure shows the total throughput of the application as a function of the size of the system, where the dashed line indicates the ideal scaling curve. The bottom figure shows the total event rate of the application as a function of the size of the system, where the dashed line indicates the system requirement of \mbox{$30\,$MHz}. \textbf{Our network scales to the full size of the event builder cluster and is capable of handling a throughput over the design rate.}

\subsection{Full system results}

\begin{figure}[hbt!]
  \centering
  \includegraphics[width=\linewidth]{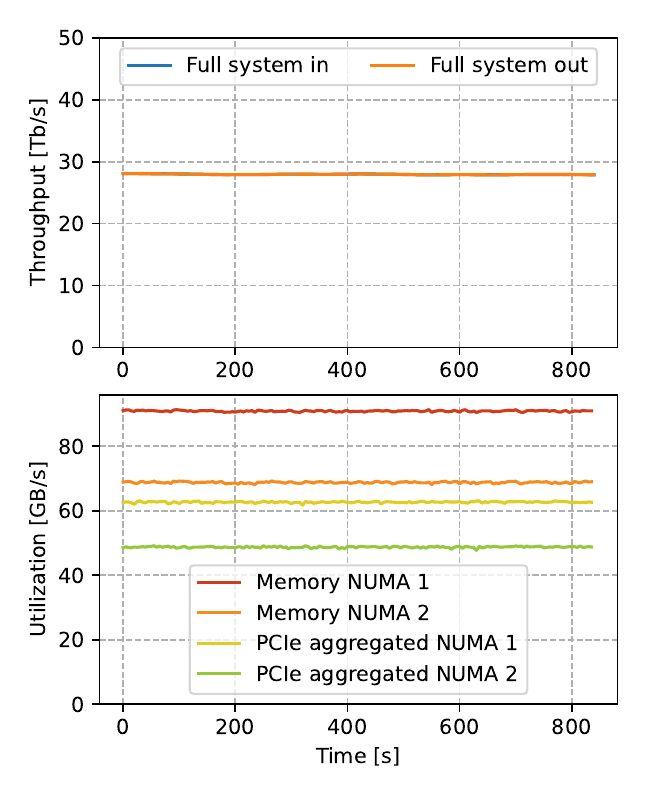}
  \caption{Full integration test of the EB+HLT1 applications. Top:~Throughput attained, corresponding to 30~MHz of physics collisions. Bottom: Memory and PCIe utilization of one node during the test.}
  \label{fig:eb_full_test}
\end{figure}

Figure~\ref{fig:eb_full_test} depicts the full integration test of the event building and HLT1 applications. The test has been conducted using a realistic event size model, and the nominal input rate of \mbox{$30\,$MHz}. The FPGAs on each node are generating data, and sending it to main memory according to the data streams depicted in Figure~\ref{fig:single_server}. In addition, a data generator is producing data, which flows through the EB~network in and out, performing the event building, and distributing the data to the GPUs. Each GPU copies the required data and performs a passthrough selection with an acceptance rate of 30:1, sending back to CPU memory only those events that passed the selection. This test replicates the expected data flow of our entire application. The data generator produces additional memory pressure that does not affect the performance of our application.

The top of Figure~\ref{fig:eb_full_test} shows the aggregated throughput of the data flowing through the EB application. The bottom Figure depicts the aggregated memory and PCIe throughput for both NUMA domains of one server during the tests. NUMA domain~1 reports a higher PCIe aggregated utilization, which is due to the asymmetric setup of FPGAs. There is a noticeable imbalance in memory bandwidth utilization between NUMA domains due to the data generation process that was run as part of these tests. Each node is sustaining a memory utilization of 160~GB/s and an aggregated PCIe~utilization of 115~GB/s. \textbf{The full system achieves the target throughput with no performance drops.}

\section{Implications}

This work is the culmination of a decade of R\&D work in preparation for the upgrade of LHCb and the restart of data taking. It has led to the highest throughput processed in software in real-time of any physics experiment to date, going beyond and surpassing its design requirements enabling a wider physics programme in the experiment. The converged architecture presented with this work will be instrumental in pushing the boundaries of our fundamental physics understanding in the following years.

\subsection{Data acquisition systems}

Historically High Energy Physics data acquisition systems have started out as mostly custom-made electronics with a strong R\&D focus on electronics engineering. Over the years, as computing and network hardware evolved, commercial-off-the-shelf hardware has more and more found its way into our acquisition systems~\cite{Colombo2018}. With our current work we continue this ongoing effort. With the radical re-design of our readout network and hardware, we managed to move away from expensive VME and ATCA hardware and consolidate everything into a PCI~Express format which has drastically increased the accessibility to compatible, mass produced, commercial hardware. It has also allowed us to forgo the deep buffer, high-end core routers for more affordable, close-to-zero buffer top of rack and director switches.

By moving closer toward the HPC model, we are profiting from this quickly growing community. At the same time we are also giving back to that community by presenting hardware manufacturers with our rather uncommon use case. In particular we decided to go with an InfiniBand interconnect, which is mainly used in HPC for its extremely low latency data transport first and bandwidth second. In the beginning of our R\&D efforts it was not obvious how to run an IB network of this size at such high link loads in an all-to-all pattern. Working in close cooperation with the hardware manufacturers we managed to steer the technology to a point where this is now possible.

\subsection{Software for real-time data processing}

The removal of the low-level trigger is a long sought goal of any physics experiment, as it reduces bias in event selection by considering the complete event instead of portions of it. We have proven that we can take data at the design rate and process it in real-time. The trend for GPU adoption to offload specialized work in physics triggers has been present in the field during the last years. We have taken the next step by processing HLT1 completely on GPUs, pushing the state-of-the-art in physics reconstruction.

The GPU-based framework that we have developed has outgrown its initial HLT1 focused purpose. We have realized that we can put more work than was initially intended, and several algorithms are currently in development that significantly improve the efficiency of targeted physics analyses. Our work is also a stepping stone for the foreseen LHCb upgrades~\cite{LHCbCollaboration:2776420,Aaij:2636441}, and demonstrators of the feasibility of a fully GPU-based trigger are under preparation.

Our software is having an impact on the LHCb detector evolution. The effect new detector technologies have is encompassed by software studies that empirically test changes in physics efficiency and throughput. This feedback loop also exists in hardware. Opportunistic preprocessing is being developed in the existing FPGA acquisition boards in tandem with the high-level trigger, potentially reducing software memory pressure. LHCb is taking a holistic approach to the optimization of its architecture.

At the same time, making our trigger a purely high-level trigger has allowed us to draw from a vast pool of open source tools and other software allowing us to develop faster and giving us more flexibility. It also opened up the software development process to a larger audience, making it more accessible, which in turn improves quality.

\section{Acknowledgements}

The authors would like to acknowledge the support of the LHCb collaboration throughout the development of the presented results. We thank the LHCb Online team for the hardware support during our tests. We would also like to thank the LHCb computing, RTA and simulation teams for their support and for producing the simulated LHCb samples used to develop and benchmark our algorithm. We would like to thank Ben~Couturier and Xavier~Vilas\'is~Cardona for useful discussions which improved this paper. We thank Chris~Broekema for his insights on radio astronomy applications. We thank David~Rohr for his valuable input in the elaboration of the comparison table. We thank NVIDIA and especially Andreas~Hehn for the support and fruitful discussions.

Affiliations of all authors
Roel~Aaij:
Nikhef National Institute for Subatomic Physics, Amsterdam, Netherlands;
Thomas~Boettcher:
University of Cincinnati, Cincinnati, OH, United States
Christina~Agapopoulou, Daniel~Hugo~C\'ampora~P\'erez, Tommaso~Colombo, Flavio~Pisani, Rainer~Schwemmer, Niko~Neufeld, Alberto~Perro, Saverio~Mariani, Arthur~Hennequin, Marian~Stahl, Rosen~Matev, Renato~Quagliani:
European Organization for Nuclear Research (CERN), Geneva, Switzerland;
Dorothea~vom~Bruch:
Aix Marseille Univ, CNRS/IN2P3, CPPM, Marseille, France;
Adrian~Casais~Vidal:
Instituto Galego de F\'isica de Altas Enerx\'ias (IGFAE), Universidade de Santiago de Compostela, Santiago de Compostela, Spain;
Daniel~C.~Craik:
Physik-Institut, Universität Zürich, Zürich, Switzerland;
Niklas~Nolte, Kate~A.~Richardson:
Massachusetts Institute of Technology, Cambridge, MA, United States;
Placido~Fernandez~Declara:
Department of Computer Science and Engineering, University Carlos III of Madrid, Madrid, Spain;
Louis~Henry:
Institute of Physics, Ecole Polytechnique F{\'e}d{\'e}rale de Lausanne (EPFL), Lausanne, Switzerland;
Patrick~Spradlin:
School of Physics and Astronomy, University of Glasgow, Glasgow, United Kingdom;
Marianna~Fontana, Alessandro~Scarabotto, Vladimir~V.~Gligorov, Christina~Agapopoulou,:
LPNHE, Sorbonne Universit\'{e}, Paris Diderot Sorbonne Paris Cit\'{e}, CNRS/IN2P3, Paris, France;
Tim~Evans, Florian~Reiss:
Department of Physics and Astronomy, University of Manchester, Manchester, United Kingdom;
Arantza~Oyanguren, Brij~Kishor~Jashal, Jiahui~Zhuo:
IFIC (Instituto de F\'isica Corpuscular), University of Valencia-CSIC, Valencia, Spain;
Marianna~Fontana is now affiliated with Sezione INFN di Bologna, Bologna, Italy;
Tim~Evans is now affiliated with Nikhef National Institute for Subatomic Physics, Amsterdam, Netherlands;
Christina~Agapopoulou is now affiliated with IJCLab - Laboratoire de physique des 2 infinis Irene Joliot-Curie, Inst. Nat. Phys. Nucl. \& Particules (CNRS/IN2P3), Paris, France;
Adrian~Casais~Vidal is now affiliated with Massachusetts Institute of Technology, Cambridge, MA, United States;
Marian~Stahl is now affiliated with  Institute of Experimental Physics I, Ruhr-Universitaet Bochum, Germany;
Alessandro~Scarabotto is now affiliated with the Technische Universit{\"a}t (TU) in Dortmund;
Alberto~Perro is also affiliated with Università degli studi di Torino, Torino, Italy, and Sezione INFN di Torino, Torino, Italy;
Florian~Reiss is now affiliated with Physikalisches Institut, Universitat Freiburg, Germany;

Alessandro~Scarabotto, Christina Agapopoulou, Marianna Fontana and Vladimir~V.~Gligorov were supported by the European Research Council under Grant Agreement number 724777 ``RECEPT''. Dorothea~vom~Bruch acknowledges support of the European Research Council Starting grant ALPACA 101040710. Niklas~Nolte was supported by NSF Grant OAC-2004645. Kate~A.~Richardson was supported by NSF Grant OAC-1836650 and is now supported by NSF Grants OAC-1836650 and PHY-1904160. Tim~Evans was supported by ERC Starting Grant 852642 BEAUTY2CHARM. Louis~Henry acknowledges the support from the SPARK funding scheme.


\bibliographystyle{IEEEtran}
\bibliography{IEEEabrv,libsettings,library,zotero_library,custom_library}

\begin{IEEEbiography}[{\includegraphics[width=1in,height=1.25in,clip,keepaspectratio]{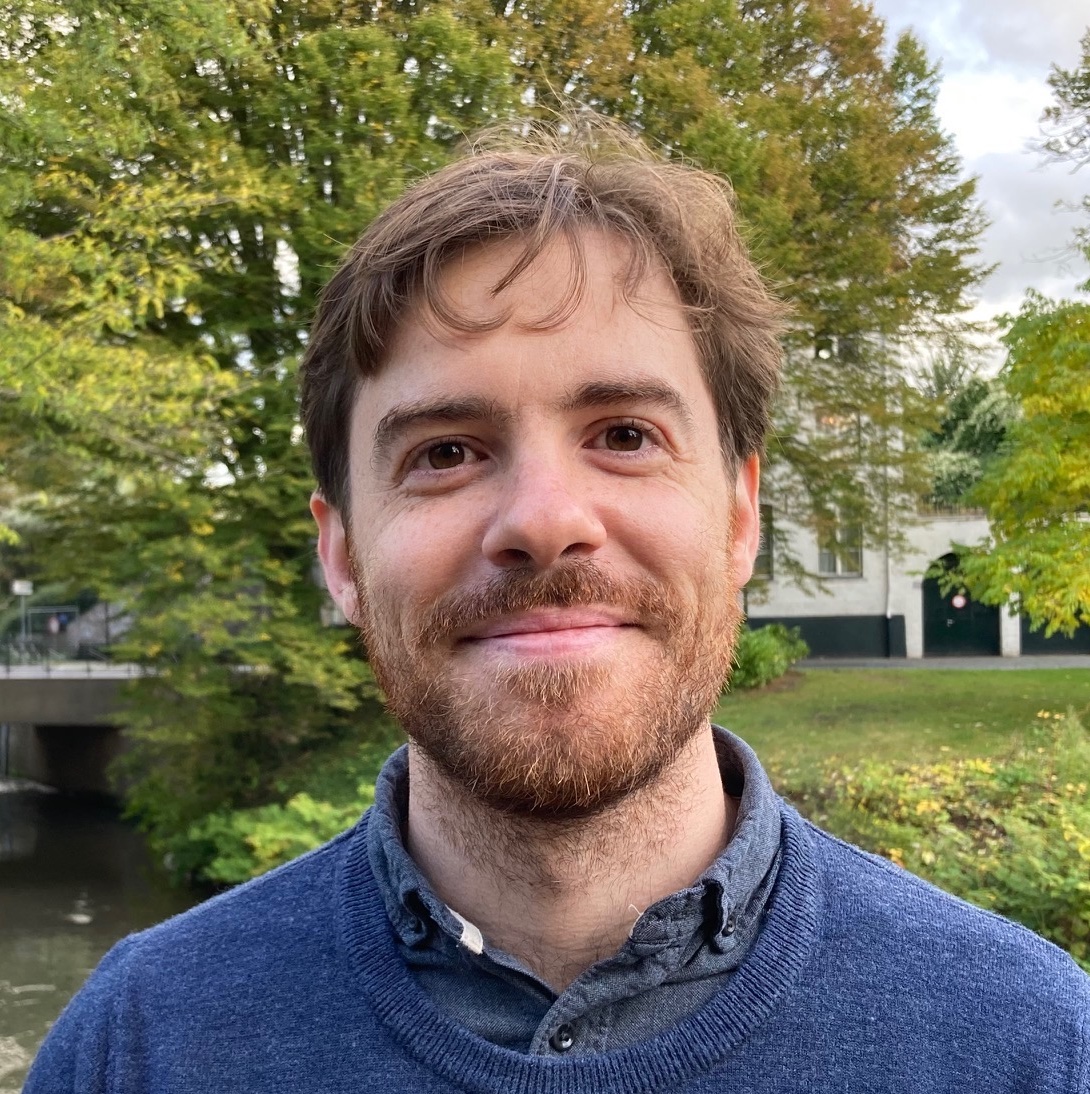}}]{Daniel Hugo Cámpora Pérez} received his Ph.D. from the University of Sevilla. He has worked at CERN since 2010, contributing mainly to the Online and software recontruction of the LHCb experiment. His main focus is in the optimization of high-throughput real-time processes in physics reconstruction with parallel architectures. In 2023 he joined NVIDIA and he is working since on parallelization of A.I. in the TensorRT-LLM framework.
\end{IEEEbiography}

\begin{IEEEbiography}[{\includegraphics[width=1in,height=1.25in,clip,keepaspectratio]{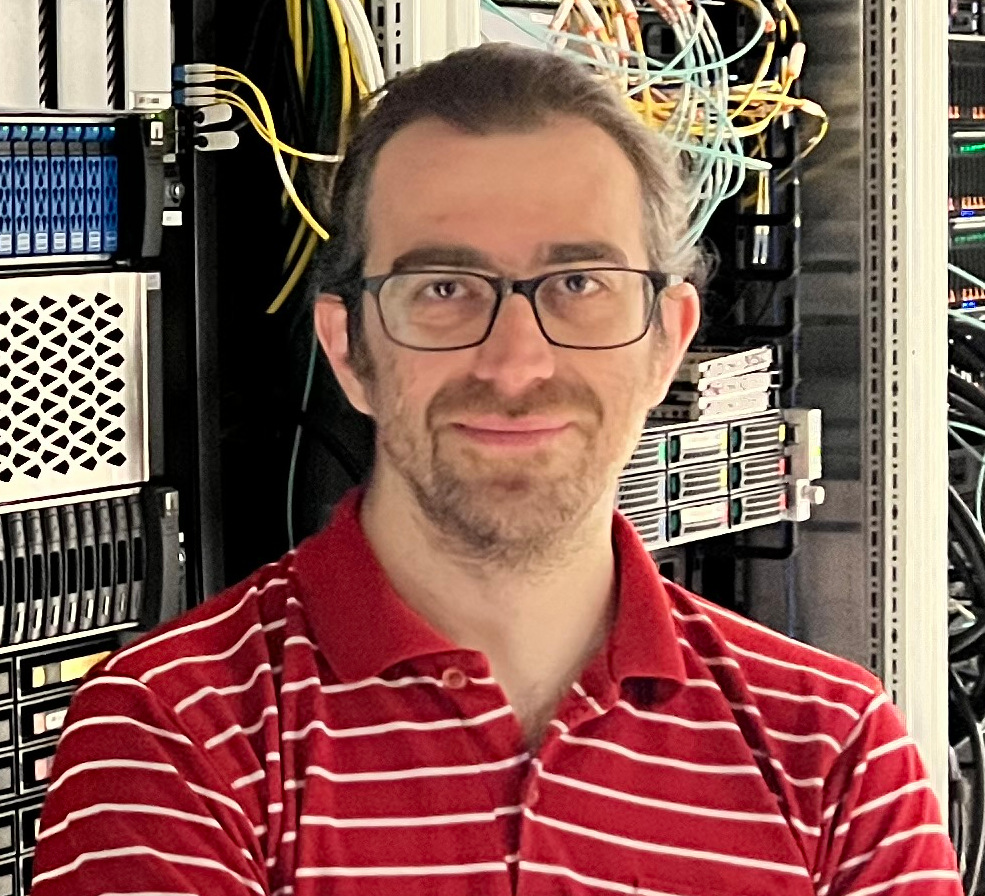}}]{Flavio Pisani} is a staff member at CERN. He received his Ph.D. degree in physics from the University of Bologna "Alma Mater Studiorum''. He is a physicist specialized in data acquisition systems and high-throughput interconnection technologies. Currently, he is responsible for the DAQ network and software of the LHCb experiment.
\end{IEEEbiography}

\begin{IEEEbiography}[{\includegraphics[width=1in,height=1.25in,clip,keepaspectratio]{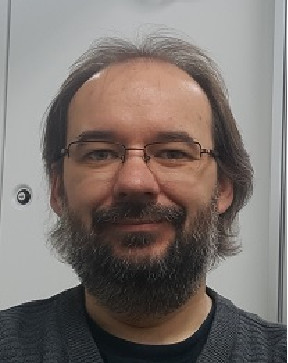}}]{Rainer Schwemmer} has graduated as a physicist from the University of Heidelberg, specializing in the development and construction of detectors for particle physics experiments. Since 2011 he is a CERN Staff Scientist working on all aspects related to large scale data acquisition systems. During his tenure at CERN, among other things, he has worked on Control Systems, High Speed Network and Data Storage, Systems Engineering and Software Performance Optimizations.
\end{IEEEbiography}

\EOD

\end{document}